# 2.7-octave supercontinuum generation spanning from ultraviolet to near-infrared in thin-film lithium niobate waveguides

Minghui Li,[1,5] Qiankun Li,[2] Yongyuan Chu,[2] Youting Liang,[3] Hairun Guo,[2,*] Jintian Lin,[1,5,*], Xueying Sun,[2] Hongyang Shi,[2] Xinzhi Zheng,[3,4] and Ya Cheng[1,3,4,6,7,8,9,*]

[1]*State Key Laboratory of Ultra-intense laser Science and Technology, Shanghai Institute of Optics and Fine Mechanics, Chinese Academy of Sciences, Shanghai 201800, China*
[2]*Key Laboratory of Specialty Fiber Optics and Optical Access Networks, Shanghai University, Shanghai 200444, China*
[3]*State Key Laboratory of Precision Spectroscopy, East China Normal University, Shanghai 200062, China*
[4]*The Extreme Optoelectromechanics Laboratory (XXL), School of Physics and Electronic Science, East China Normal University, Shanghai 200241, China*
[5]*Center of Materials Science and Optoelectronics Engineering, University of Chinese Academy of Sciences, Beijing 100049, China*
[6]*Shanghai Research Center for Quantum Sciences, Shanghai 201315, China*
[7]*Hefei National Laboratory, Hefei 230088, China*
[8]*Collaborative Innovation Center of Extreme Optics, Shanxi University, Taiyuan 030006, China*
[9]*Collaborative Innovation Center of Light Manipulations and Applications, Shandong Normal University, Jinan 250358, China*
E-mail: hairun.Guo@shu.edu.cn; jintianlin@siom.ac.cn; ya.cheng@siom.ac.cn;

Supercontinuum generation (SCG) with spectral coverage across the full visible and ultraviolet (UV) ranges is crucial for optical clocks, quantum computing and sensing. However, achieving such SCG in nanophotonic platforms is challenging due to the difficulties in spectrum broadening. Here, Such ultrabroad-bandwidth SCG was demonstrated in thin-film lithium niobate (TFLN) nanophotonic waveguides by dispersion management, without periodic poling for spectral broadening. Anomalous-dispersion waveguides were designed in the telecom band, simultaneously enabling dispersive wave emergence, modal-matched second harmonic generation, and third harmonic generation for spectrum broadening. Moreover, MgO was intentionally doped to mitigate the photorefractive effect of lithium niobate, which frequently results in un-sustained spectrum broadening and in turn limits the accessible SCG coverage. By leveraging photolithography assisted chemo-mechanical etching, low-loss MgO doped TFLN nanophotonic waveguides were fabricated. As a result, thanks to the utilization of the strong $\chi^{(2)}$ and $\chi^{(3)}$ nonlinear processes, gap-free 2.7-octave SCG spanning from 330 nm to 2250 nm was observed by pumping the waveguide with a 1550-nm femtosecond pulsed laser with 0.687 nJ, agreeing well with numerical simulation. This spectral coverage represents the state of the art in TFLN platforms without fine microdomains, and even close to the record in sophisticated chirped periodically poled TFLN waveguides.

## 1. Introduction

Optical frequency combs (OFCs) have demonstrated tremendous potential across a wide range of applications, including precision metrology, optical clocks, molecular spectroscopy, and biosensors, since their first demonstration in the near-infrared band two decades ago.[1-7] There is an increasing demand to extend OFC coverage across the entire optical spectrum, particularly from benchtop systems to chip-level platforms,[8-21] to enable scientific discoveries and photonic applications at compact footprint, low power consumption, and reduced costs. However, achieving robust OFC coverage across the full visible and ultraviolet (UV) ranges remains challenging on integrated platforms, despite its critical importance for numerous quantum and atomic systems.[22]

Recent advancements have shown gap-free OFC coverage spanning 330-2400 nm in thin-film lithium niobate (TFLN) platforms through supercontinuum generation (SCG) by combining $\chi^{(2)}$ and $\chi^{(3)}$ nonlinearities.[23] This achievement is enabled by the exceptional properties of TFLN photonic waveguides, which simultaneously exhibit strong $\chi^{(2)}$ nonlinearity ($d_{33}$ = 30 pm/V, $d_{31}$ = 5.9 pm/V, $d_{21}$ = -3 pm/V), a large nonlinear refractive index ($n_2$ = 2.5 × $10^{-19}$ $m^2$/W), a broad transparency window from 0.33 to 5 µm, a strong ferroelectric effect, and high optical confinement.[24-32] Notably, multisegment nanophotonic TFLN waveguides with engineered dispersion and chirped periodic poling were employed for spectrum broadening, requiring sophisticated manufacturing processes such as expensive electron-beam lithography and rigorous chirped periodic poling.[23]

Therefore, developing SCG across the full visible and UV ranges in simply structured nanophotonic waveguides while relaxing fabrication tolerances and reducing manufacturing costs remains a significant challenge. Furthermore, TFLN waveguides fabricated using inductively coupled plasma reactive ion etching (ICP-RIE) have suffered from high surface roughness, leading to large scattering losses of optical modes in previous demonstrations.[20,23,32] This limitation restricts the bandwidth of SCG when pumped at telecom wavelengths. Additionally, SCG in these waveguides undergoes periodic changes in refractive index under high-power pumping due to the photorefractive effect, affecting long-term retention of wide SCG and limiting phase locking to narrow-band spectra.[20]

In this work, we demonstrate broad-bandwidth SCG with a total bandwidth spanning 2.7 octaves (330 nm to 2250 nm) in MgO-doped TFLN ridge waveguides fabricated using photolithography-assisted chemo-mechanical etching (PLACE),[31] pumped by a 1550-nm mode-locked laser source. To achieve such broad-bandwidth SCG, dispersion engineering was performed with anomalous group velocity dispersion (GVD) at the telecom-band pump wavelength, enabling dispersive wave emergence, modal-matched second harmonic generation (SHG), and third harmonic generation (THG) for spectrum broadening. The MgO doping suppressed material damage and photorefractive effects, allowing long-term maintenance of SCG—a critical factor for achieving record-bandwidth SCG in TFLN platforms without requiring fine microdomain structures. Numerical simulations using the split-step Fourier method were conducted to support the experimental results, accurately predicting soliton fission position and spectrum profile.[32,33] For comparison, SCG was also demonstrated in undoped TFLN waveguides, showing narrower bandwidths due to increased scattering losses and photorefractive effect. These findings highlight the importance of MgO doping in enhancing the performance of TFLN-based SCG devices. This work advances the development of integrated OFC sources for applications requiring broad spectral coverage across the visible and UV ranges.

## 2. Supercontinuum generation in MgO: TFLN waveguide

### 2.1. Fabrication of nanophotonic waveguides

Commercially available Z-cut MgO: TFLN wafer consisting of a 900-nm-thick MgO: TFLN layer, a 2-μm-thick silica layer and 500-μm-thick lithium niobate (LN) handle was used as platform to produce the dispersion engineered nanophotonic waveguides. The MgO-doped concentration was chosen as 5 mol.% to significantly suppress the strong photorefractive effect of LN. The dispersion engineered waveguides were fabricated by PLACE technique.[31] The etched depth of the LN thin film was ~662 nm, and a secondary CMP process is necessary to improve the surface smoothness of the waveguides.[31,34] The fabricated waveguides oriented along X axis of LN, and possess a top width of 2 μm, a wedge angle of ~8.36º, and ultra-smooth sidewalls for suppressing scattering loss.

### 2.2. The experimental setup

The experimental setup for SCG is illustrated in Fig. 1. A mode locked laser (Model: CFL-ZCFF, CARMEL) centered at 1550 nm with a repetition rate of 80-MHz was used as pump source. A combination of half-wave plate (HWP, Thorlabs: AHWP05M-1600) and polarizing beam splitter (PBS, Thorlabs: CCM1-PBS25-1550) was used to keep the incident light to be horizontally polarized and change the light intensity arbitrarily while maintaining the pulse width at 81 fs. The reflected light was blocked by an optical trap, whilst the transmitted light was launched into the waveguide through an aspheric lens (Thorlabs: C230CMD-C) with numerical aperture of 0.14 to excite SCG in the waveguide. And the SCG signal was collected through a single-mode lensed fiber and sent to optical spectrum analyzer (OSA, Model: AQ6315A and Model: AQ6375) for spectral analysis.

## 2.3. Ultrabroad-bandwidth SCG

### 2.3.1. SCG in MgO: TFLN waveguides

Figure 2 (a) shows a supercontinuum generated in a MgO: TFLN waveguide with a top width of 2 µm and 6.5 mm length under optical pump with 0.687 nJ pulse energy. Ultrabroad-bandwidth SCG was observed with gap-free coverage ranging from 330 nm to 2250 nm, spanning 2.7 octaves of bandwidth. And the detected spectral range is limited by the OSA, which only works at the range shorter than 2250 nm. To clearly confirm the gap-free envelope, an enlarged spectrum region ranging from 330 nm to 500 nm was analyzed by a grating spectrometer which could detect small signals down to -90 dBm, as plotted in Fig. 2 (b). The red curve displays a gap free spectrum structure under 0.687-nJ pump, without considering the coupled loss (~-10 dB) of the output signals. For comparison, the black curve in Fig. 2 (b) represents discontinuous spectrum pumped by an incident pulse power of 0.625 nJ in the same waveguide. Moreover, due to the suppression of photorefractive effect in LN by doping MgO, the SCG maintains for a long time of period. The inset in Fig. 2 (b) shows a photograph of soliton fission scattering light collected above the waveguide, which is distinguishable by the green third harmonic wave and blue light emission as light propagation.

The SCG dependence on the input power was also characterized. Figure 3 shows the SCG spectra varied with pump pulse energies. At pulse energy of 0.27 nJ, the onset of modal-matched SHG at 775 nm and THG at 517 nm as well as short wavelength dispersion wave (SWDW) around 1100 nm was observed. As the pump pulse further increased, the SHG, THG and self-phase modulation signals started broadening. The SHG and SCG components began to spectrally overlap when pulse energy reached 0.405 nJ, which is necessary for f-2f reference. Sequentially, at 0.54 nJ pulse energy, both sum frequency generation (SFG) and SHG as well as self-phase modulation cover the spectrum range from 424.6 nm to 2400 nm, spanning 2.5 octaves of bandwidth. Therefore, the ultrabroad-bandwidth SCG results from the combination of excellent $\chi^{(2)}$ and $\chi^{(3)}$ nonlinearities, and the mitigation of photorefractive effect of MgO: TFLN nanophotonic waveguides.

### 2.3.2. Dispersion engineering and analysis

To reveal the underlying mechanics behind ultrabroad-bandwidth SCG, it is necessary to analyze the dispersion of the nanophotonic waveguides. Figure 4 (b) depicts the simulated dispersion curve considering the cross section of the waveguide illustrated in Fig. 4 (a), where the red dashed line shows an anomalous dispersion at 1550 pump wavelength. The zero-point of integrated dispersion $\beta_{int}$ (green line) nearly 1100 nm represents the position of short wavelength dispersive wave (SWDW), enabling to flatten the spectrum and further broaden it to near UV band.

### 2.3.3 Simulation and analysis of the spectra

The theoretically simulated SCG spectrum in the MgO: TFLN waveguide with 6.5 mm length is displayed in Fig. 4 (c), which was conducted by solving generalized nonlinear Schrödinger equation (GNLSE) through spit-step Fourier method,[33] with a pump energy of 0.81 nJ of $TE_{00}$ mode. Altogether, the simulated result agrees well with experimentally measured spectrum displayed in Fig. 2 (a) at 300 nm to 2400 nm in terms of profile. Detailed compared with simulated result, the experimental spectrum shows strong discrete spectrum component nearly 500 nm, thus enhanced comb power at this location and extended the spectrum to the UV band. Figures 4 (d) and 4 (e) show temporal and frequency domain evolution of pump pulse traveling along the waveguide, respectively, exhibiting self-phase modulation, soliton fission and dispersive wave (DW) generation at both domain as pulse transmission.

## 3. Discussion and Conclusion

### 3.1 SCG in undoped LN waveguides

For comparison, SCG in an undoped Z-cut TFLN waveguide with 600 nm thickness, as its cross section illustrated in the inset of Fig. 5 was collected. The cross section of the waveguide was designed to maximize the spectrum broadening, which differs from the MgO: TFLN counterpart because of different refractive indices. The integrated dispersion of the waveguide is shown by the green curve in Fig. 5, exhibiting a SWDW at 969 nm. To experimentally demonstrate SCG in the undoped waveguide, the waveguide with one input optical spot size converter was used to fabricated by PLACE technique to improve the coupling efficiency to nearly 11.3 dB.[35] The experimentally collected SCG is shown by the black curve in Fig. 5, corresponding to 1.636 octaves, pumping at 0.373 nJ by $TE_{00}$ mode. Its bandwidth is limited by the less mode confinement of the film thickness through light scattering. During the experiment, due to photorefractive effect of LN, the ultrabroad bandwidth SCG does not maintain for a long time. Meanwhile, the bandwidth of SC is narrow, consistent with the previously reported results.[15,20]

In the future, the coupled loss should be reduced for increasing the collected output power of SCG for f-2f reference. Alternatively, through changing the wavelength of pump source, we could further achieve f-2f beat detection and mode locking while maintaining ultrabroad SCG, by leveraging the mitigation of photorefractive effect of LN.[20]

**3.2 Conclusion**

To summarize, a 2.7-octave SCG spanning from UV to infrared band was demonstrated with a gap-free envelope in the integrated MgO: TFLN waveguides fabricated by PLACE technique, without introducing chirped periodically poled structures. Our results pave the way for realization of lower cost chip-scale optical clocks and higher resolution optical tomography based on TFLN platforms. [4,36,37]

**Acknowledgements**

We thank Prof. Jiaqi Zhou at Shanghai Institute of Optics and Fine Mechanics for providing one mode-locked fiber laser.

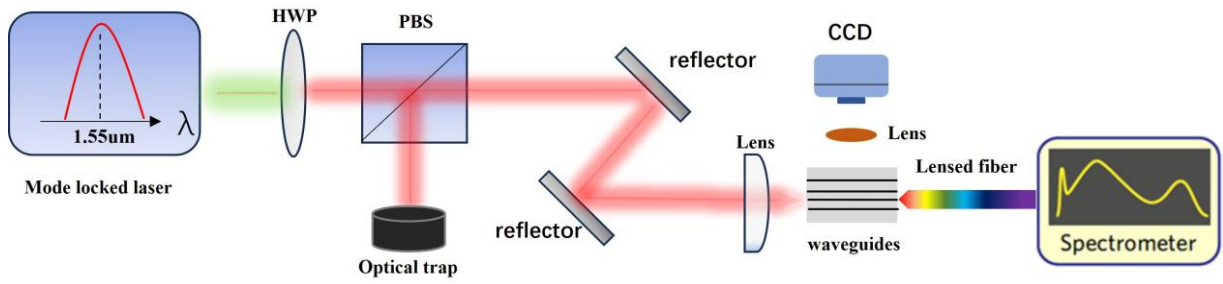

**Figure 1**. Experimental setup for SCG. HWP: Achromatic half wave plate; PBS: Polarizing beam splitter; CCD: Charge coupled device.

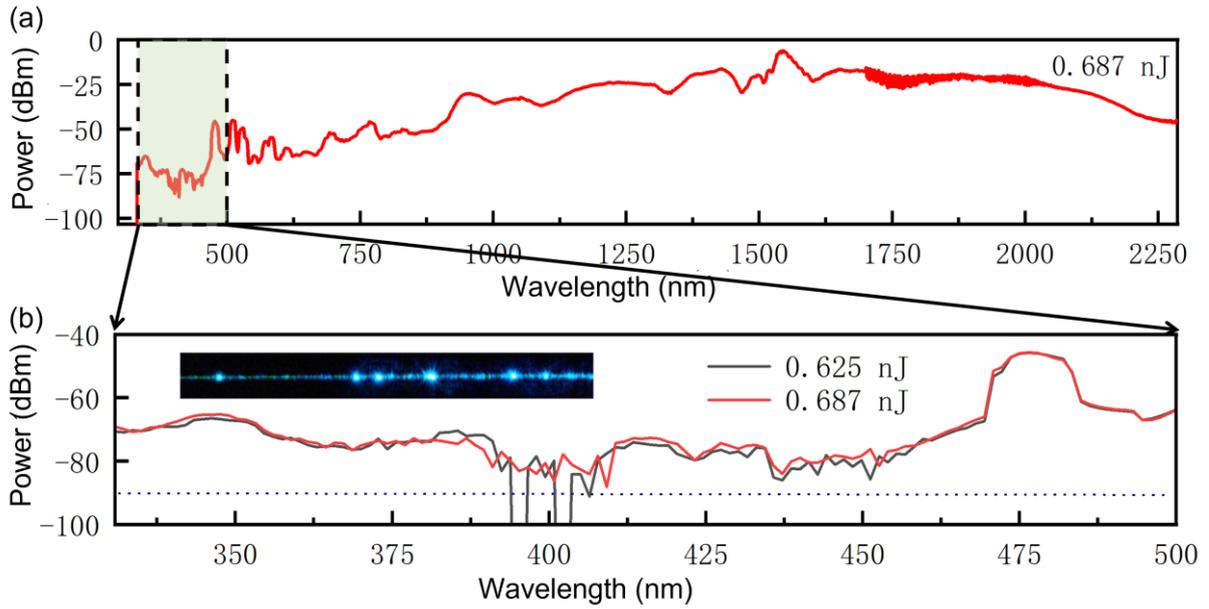

**Figure 2**. (a). Spectrum of 2.7-octave SCG in the MgO: TFLN waveguide pumped by $TE_{00}$ mode at 0.687-nJ pump level. (b). Enlarged spectrum regions ranging from 330 nm to 500 nm of Fig. 2 (a) (red curve) and another SCG spectrum generated in the same waveguide (black curve) at a low pump level of 0.625 nJ for comparison, confirming a gap free spectral envelope of SCG in Fig. 2(a). And the lower detected limit of the spectrometer is labeled with blue dotted line. Inset: Photograph taken above the waveguide, showing bright scattering light emission of the soliton fission. The captured length is ~0.6 mm.

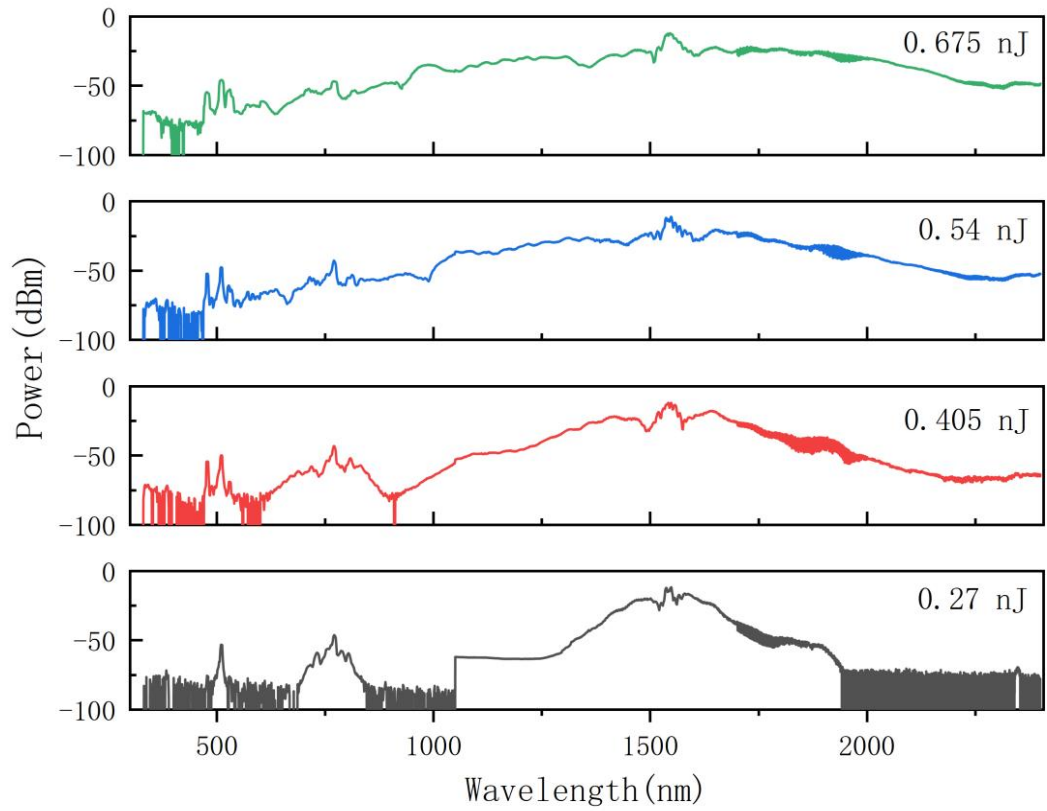

**Figure 3.** Measured spectra at various pump pulse energies of 0.27 nJ, 0.405 nJ, 0.54 nJ, and 0.675 nJ (from bottom to top).

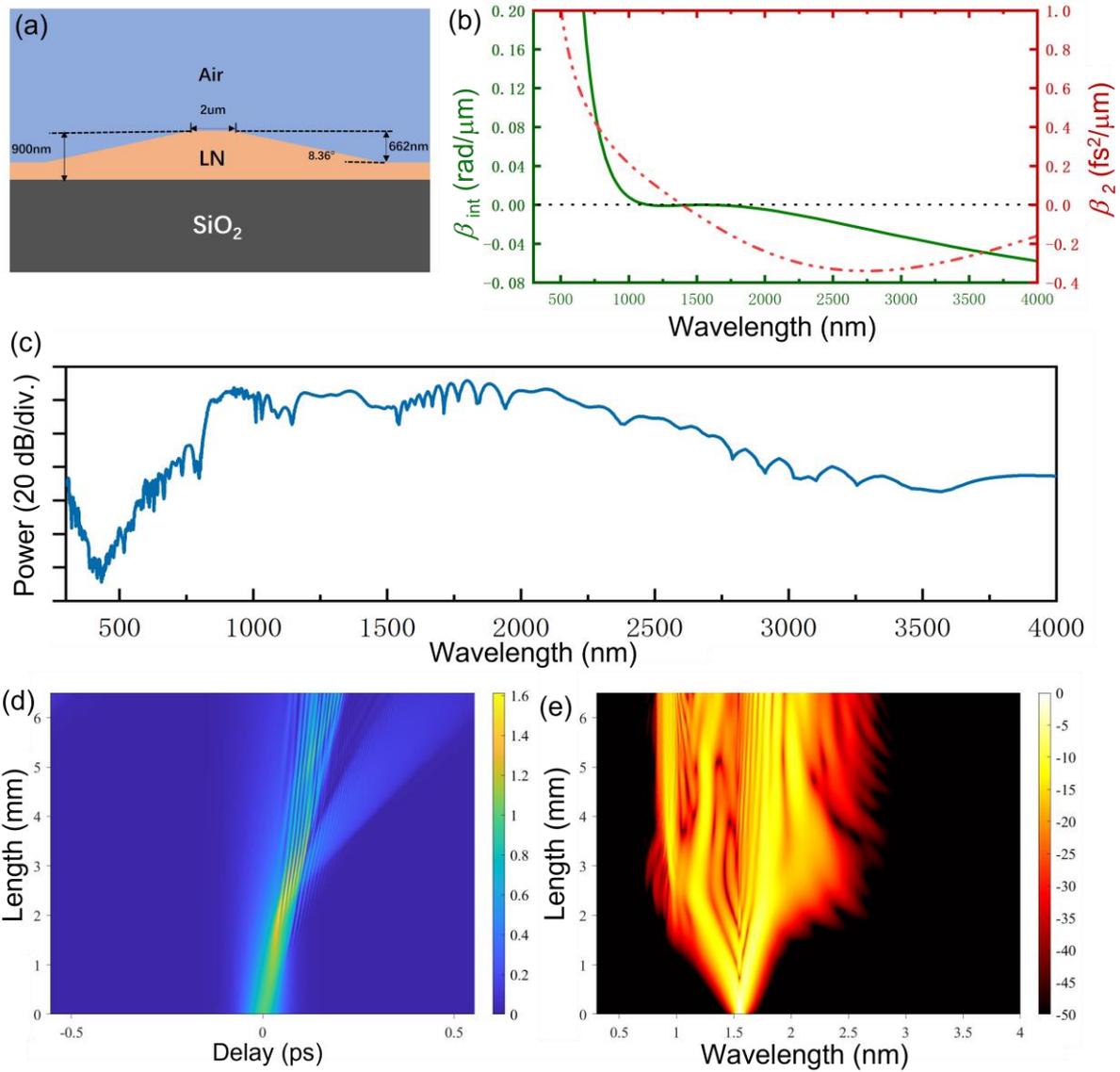

**Figure 4**. (a). Illustration of cross section of the MgO: TFLN waveguide. (b). Simulated curves of integrated dispersion $\beta_{int}$ (green solid line) and second order dispersion $\beta_2$ (red dashed line) of the waveguide. (c). Simulated spectrum of waveguide configuration shown in Fig. 4 (a) with a length of 6.5 mm. (d). Simulated temporal and e) spectral evolution in the same waveguide.

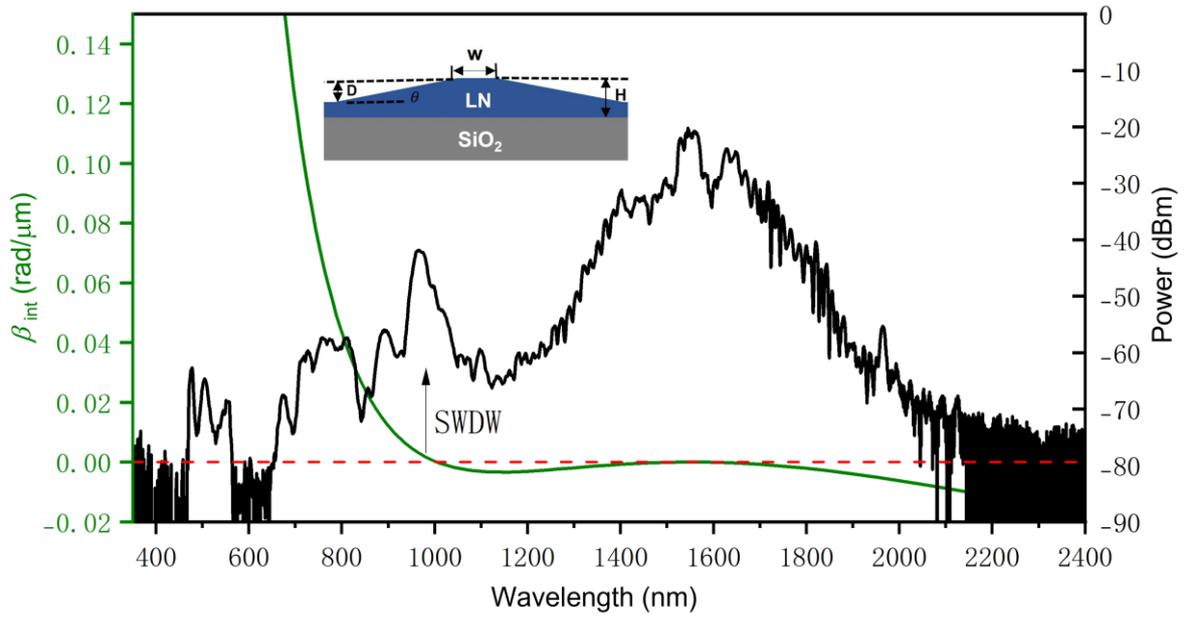

**Figure 5.** SCG in another TFLN waveguide without doping of MgO (black line) and simulated integrated dispersion $\beta_{int}$ (green line). Inset: Illustration of cross section of the undoped TFLN waveguide. Here, H (TFLN thickness) is 600 nm, W (top width) is 2 μm, D (etched depth) is 446.8 nm, $\theta$ (wedge angle of the waveguide) is 5.654°.